\documentclass[runningheads]{llncs}
\usepackage{makeidx}  
\usepackage[T1]{fontenc}
\usepackage[scaled=0.86]{helvet}
\usepackage{graphicx}
\usepackage{booktabs}
\usepackage{amsmath,amssymb}
\usepackage[table]{xcolor}
\usepackage{hyperref}
\usepackage{tcolorbox}
\usepackage{multirow}
\usepackage{comment}
\usepackage{colortbl}
\usepackage{cite}
\usepackage{orcidlink}
\usepackage{doi}
\usepackage{xspace}
\usepackage{tabularx}

\newcommand{\PUCM}{\textsf{PM4Py-UCM\xspace}}
\newcommand{\PMPy}{\textsf{PM4Py\xspace}}
\newcommand{\jUCMNav}{\textsf{jUCMNav\xspace}}
\newcommand{\tax}[2]{\textbf{\texttt{#1}}: #2}

\usepackage{tikz}
\definecolor{MyPurple}{RGB}{112,48,160}
\newcommand{\coloredcircle}[2]{%
  \tikz[baseline=(char.base)]{
    \node[shape=circle, draw=none, fill=#1, inner sep=1pt, minimum size=7pt] (char)
    {\textcolor{white}{\sffamily\bfseries\scriptsize #2}};
  }%
}
\newcommand{\newbadge}{%
  \tikz[baseline=(char.base)]
  \node[anchor=base, fill=MyPurple, text=white, rounded corners=2pt, font=\sffamily\bfseries\tiny, inner sep=2pt] (char) {NEW};%
}

\newcommand{\contrib}[1]{\coloredcircle{MyPurple}{#1}}
\newcommand{\result}[1]{\textbf{\textcolor{MyPurple!120}{#1}}}

\newtcbox{\myinlinebox}[1][]{%
  on line,
  arc=2pt,
  outer arc=3pt,
  colback=MyPurple!10,
  colframe=MyPurple!120,
  boxsep=0pt,
  left=3pt,
  right=3pt,
  top=2pt,
  bottom=2pt,
  boxrule=0.5pt,
  #1
}

\newcommand{\lesson}[1]{\myinlinebox{#1}}

\begin{document}
\mainmatter              
\title{Building a Process-Modeling Tool using Agentic AI: An Experience Report on \textsc{PM4Py-UCM}}
\titlerunning{Building a Process-Modeling Tool using AI}  
%
\author{Daniel Amyot~\orcidlink{0000-0003-2414-1791}}
\authorrunning{D. Amyot}   
%
\tocauthor{Daniel Amyot}
\institute{University of Ottawa, Ottawa, Canada\\
\email{damyot@uottawa.ca}
}

\maketitle              

\begin{abstract}        
Enterprise-modeling (EM) tools are often complex and hard to extend. Yet, users may want to explore new EM features and capabilities that currently do not exist. AI coding agents can help here by enabling the development of new capabilities and entire tools, but whether we can trust a modeling-language tool an LLM largely wrote remains a question. This paper reports on the AI-assisted construction of \PUCM, an open-source tool that mines Use Case Map (UCM) models from event logs. \PUCM's capabilities include some expected from process mining tools (e.g., performance heat-maps and dashboards) and distinctive ones (e.g., mined executable scenarios/variants, and model decomposition). We mined the \textit{development record itself}, composed of 18 agent sessions (374 human turns and 10{,}328 tool actions over 65 hours), 151 commits, 20 releases, and a test suite grown from 108 to 691 test functions, in order to characterize, in a single in-depth case, how the tool was built with an agent (Claude Code), complemented by an independent static assessment of the resulting code (coverage, complexity, maintainability, security, architecture). We contribute a reproducible, privacy-preserving toolkit and taxonomy that classify human turns and flag cross-cutting consistency work, agent corrections, and retracted requests. Up to version 0.7.4, fixes outnumber features 2.3:1, with ${\approx}18\%$ of turns for correcting agent errors. Feature waves dragged a measurable tail of documentation/test/notebook consistency work, and tests grew lockstep with features. We finally present lessons learned, centered on making model transformations mechanically checkable, and the oracle-based validation strategy that closed the ``the agent said it works'' gap, for responsibly engineering EM tooling with AI.


\end{abstract}

\section{Introduction}\label{sec:intro}
Graphical models have supported enterprise/process modeling and requirements engineering for over a century~\cite{laue2022-EMISAJ}, and for most of that time they were drawn by hand. Process mining (PM)~\cite{vdaalst2016} is a particular type of enterprise modeling (EM) approach that enables \textit{discovering} such models from event logs, turning recorded behavior into models represented with Petri nets~\cite{PetriNets}, BPMN~\cite{bpmn2014}, or (in our case) Use Case Maps~\cite{buhr1998,urn2018}. Yet the tools that make this possible are themselves a bottleneck: EM tooling, especially commercial solutions, is typically complex and hard to extend~\cite{emreview,adoxx2013}, limiting how quickly the community can
put new modeling and analysis ideas into practice.

The recent rise of artificial intelligence (AI) agents for coding and other long/complex tasks typically performed by humans changes this context radically~\cite{sergeyuk2025-AI-coding,bhati2026agenticAI}. This paper explores the development of process-mining tooling using a commercial coding agent based on Large Language Models (LLM), namely \textit{Claude Code}\footnote{\url{https://code.claude.com/docs/en/overview}}. This new tool was motivated by the needs of ongoing research projects with Ottawa-based hospitals, where online PM tools (which represent the majority of high-end commercial solutions~\cite{loyola2023PM}) cannot be used due to strict privacy policies preventing the storage of patient-related information on the cloud, and where locally-installed solutions do not address all the needs of these projects. Such an endeavor poses a practical research question (RQ) the EM community will increasingly face: 

\begin{tcolorbox}[colback=MyPurple!5!white,colframe=MyPurple!40!black,left=2pt,right=2pt,top=0pt,bottom=0pt,halign=center]
\textbf{RQ}: Can we trust a modeling tool an Agentic AI system largely wrote?
\end{tcolorbox}

This paper reports on \PUCM, an open-source extension of the \PMPy{} library~\cite{pm4py2023} that mines executable Use Case Map models from event logs and exports them to the \jUCMNav{} tool, the main UCM modeling environment~\cite{jucmnav2006}. As an EM-language tool, \PUCM{} encodes a standardized metamodel~\cite{urn2018}, performs a chain of model transformations (process tree $\rightarrow$ UCM model $\rightarrow$ \jUCMNav{} interchange file, and a round trip back), and must stay faithful to a reference editor (\jUCMNav). Over roughly ten weeks, it was built almost entirely through an AI coding agent, and the entire development is recorded in agent-session transcripts and version control. We exploit that record to measure how the tool was actually built. This paper provides the following contributions:

\begin{itemize}
  \item \contrib{C1} A new version of \PUCM{} that extends previous versions~\cite{pm4pyucmrew,pm4pyucmsam}) with new capabilities that include i)~filtering and activity renaming, ii)~performance overlays, iii)~discovering and comparing model families, iv)~user-defined dashboards, v)~self-contained interactive HTML reports with navigable SVG models, vi)~project saving/sharing/resuming, and vii)~executable-code export (Sect.~\ref{sec:tool}).
 
  \item \contrib{C2} A measured, reproducible account of AI-assisted PM-tool development, including human/agent interaction mix, the balance of features vs.\ fixes vs.\ corrections, a developer-identified cross-cutting consistency effort, effort per modeling concern, and how tests grew with features (Sect.~\ref{sec:results}).
  
  \item \contrib{C3} An independent, tool-based assessment of the resulting artifacts (test coverage, code complexity, maintainability, security, and architecture) corroborating the internal, oracle-based validation (Sect.~\ref{sec:quality}).
  
  \item \contrib{C4} A privacy-preserving, reusable \textit{reflexive-mining} method and toolkit, plus distilled lessons for engineering PM/EM tooling with agentic AI (Sects.~\ref{sec:method},~\ref{sec:discussion}).
\end{itemize}

Note that this is an \textit{exploratory single-case study}; we scope quantitative claims to this one project throughout, and treat the patterns as hypotheses for replication. In addition to the above sections, this paper discusses necessary background (Sect.~\ref{sec:background}), threats to validity (Sect.~\ref{sec:threats}), related work (Sect.~\ref{sec:related}), and, finally, conclusions (Sect.~\ref{sec:conclusion}).

\section{Background}\label{sec:background}
The User Requirements Notation (URN) is an international standard (ITU-T Z.151~\cite{urn2018}) that combines the Goal-oriented Requirement Language (GRL) with a process notation, namely Use Case Maps (UCM)~\cite{buhr1998}. A UCM is a graph of \textit{path nodes}, including start/end points, responsibilities (or activities in BPMN terms), OR- and AND-forks/joins for choice and concurrency, and \textit{stubs} containing plug-in sub-maps. These path nodes are optionally bound to a two-dimensional hierarchy of \textit{components}, which describe activity performers (roles, persons, systems, etc.). UCM offers an executable scenario semantics, enabling validation as well as sequence-chart and test generation. The open-source Eclipse tool \jUCMNav~\cite{jucmnav2006}\footnote{Available at \url{https://github.com/JUCMNAV/jUCMNavPlus}} is the de-facto URN editor and defines the \texttt{.jucm} (XMI) file format \PUCM{} targets. URN has two decades of applied experience across combined goal/process modeling~\cite{urnexp2022}.

Process mining discovers models from event logs~\cite{vdaalst2016}. Among existing PM algorithms, the inductive miner yields a block-structured process tree~\cite{im2013} that maps cleanly to BPMN and UCM. \PMPy~\cite{pm4py2023} is the dominant open-source Python library for PM, and represents the start point for extension with \PUCM. Prior work introduced the UCM discovery pipeline, hierarchical process decomposition, and performer binding~\cite{pm4pyucmrew}, as well as concurrency-aware variant clustering with executable scenario synthesis~\cite{pm4pyucmsam}. This paper treats those as given and focuses on the tool's broader capabilities and, centrally, on \textit{how it was built}.

Contemporary LLM-based coding agents operate largely autonomously over a repository with tools to read, edit, run, and test code, driven by natural-language instructions. Their use is spreading in practice faster than it is being studied~\cite{cui2026effects,llm4seslr}, particularly for specialized software such as modeling tooling. In a recent study, Bhati~\cite{bhati2026agenticAI} argues that development based on coding agents has shifted from code generation to delegated execution under human supervision. Here, we indeed treat the agent as the primary implementer and the human as specifier, reviewer, and domain authority, and we measure that division of labor.

\section{\PUCM{} Capabilities}
\label{sec:tool}
\PUCM{}\footnote{Code and documentation available: \url{https://github.com/ProcessMining-uOttawa/pm4py-ucm/}; Web demo: \url{https://pm4py-ucm.streamlit.app/}} extends \PMPy~\cite{pm4py2023} so that UCMs become a first-class output of process discovery. Its pipeline reuses the inductive miner~\cite{im2013} to obtain a process tree, converts it into a Python UCM object model mirroring the URN metamodel, renders it in UCM or BPMN style (with components), and serializes it to a round-trip-safe \texttt{.jucm} file for \jUCMNav. The tool is available both through a Web application (using Streamlit~\cite{streamlit2025}) and programmatically (Python API). 

\begin{figure}[t]\centering
  \includegraphics[width=\linewidth]{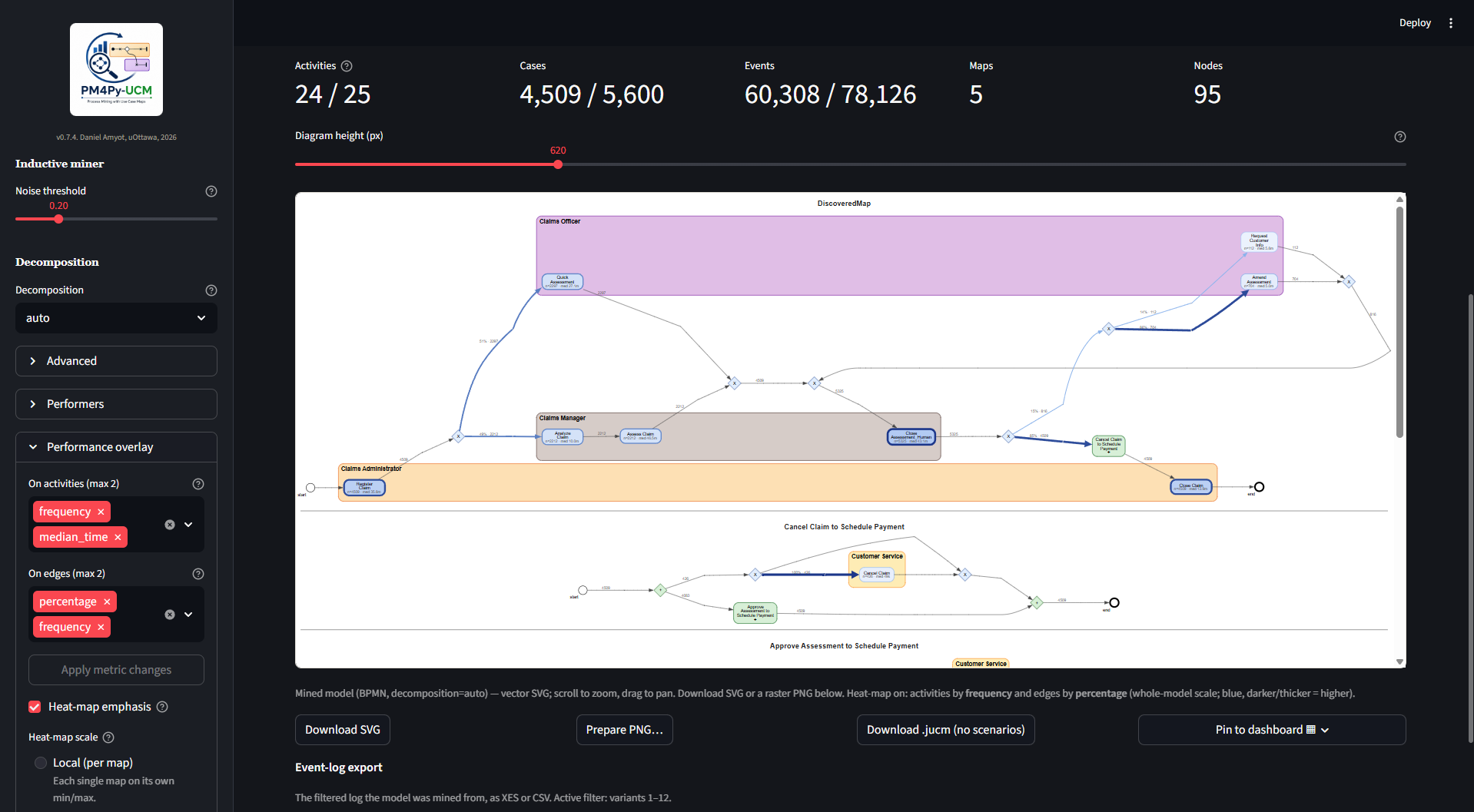}
  \caption{\PUCM{}: Streamlit-based Web interface with BPMN rendering, including performance heat-map overlays (activity/edge text with color and thickness encoding time and frequency metrics).}
  \label{fig:bpmn}
\end{figure}

\begin{figure}[t]\centering
  \includegraphics[width=\linewidth]{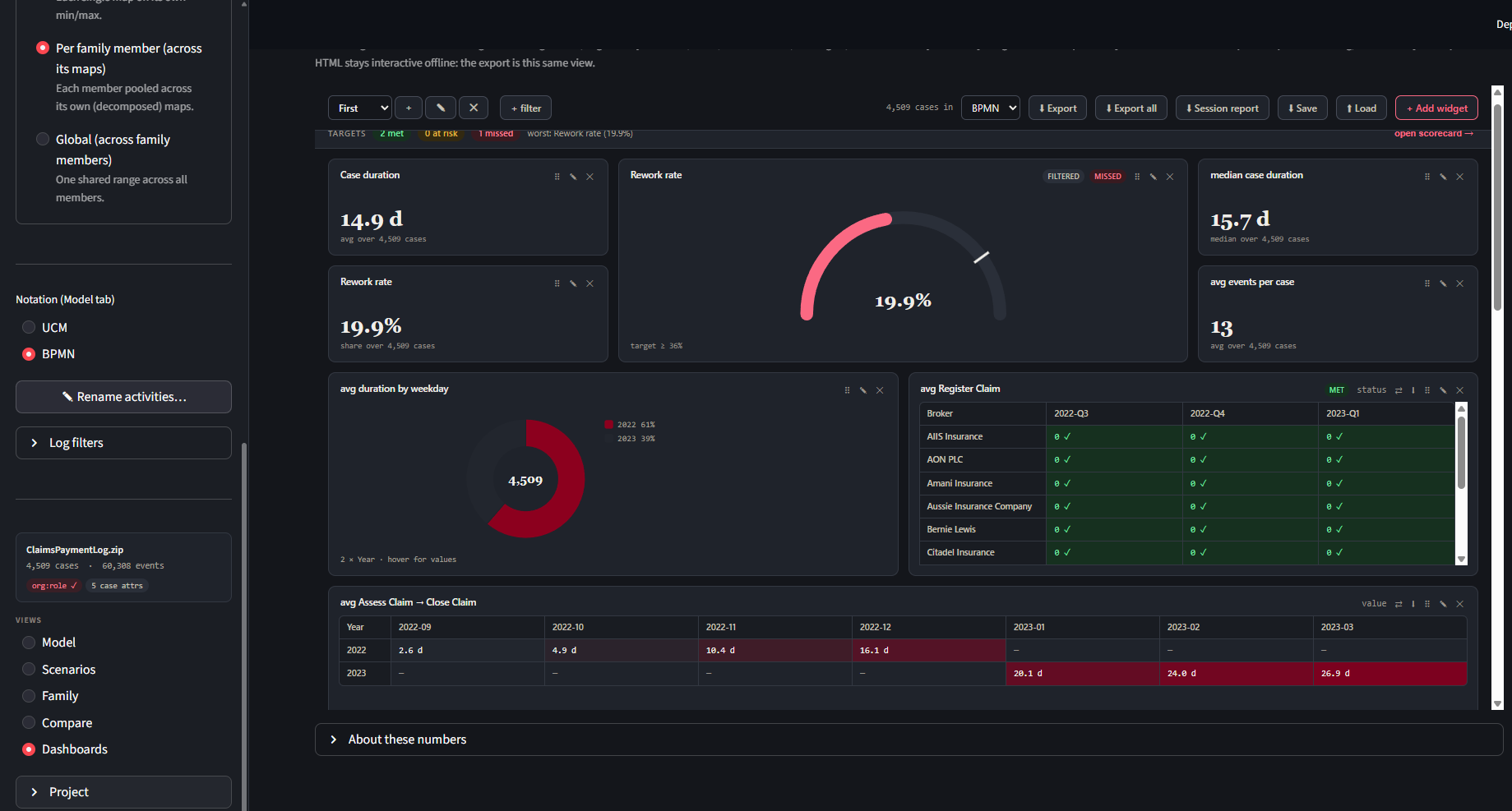}
  \caption{\PUCM{} custom analytics dashboard (KPI, gauges, distributions...).}
  \label{fig:dashboard}
\end{figure}

\begin{figure}[h!]\centering
  \includegraphics[width=\linewidth]{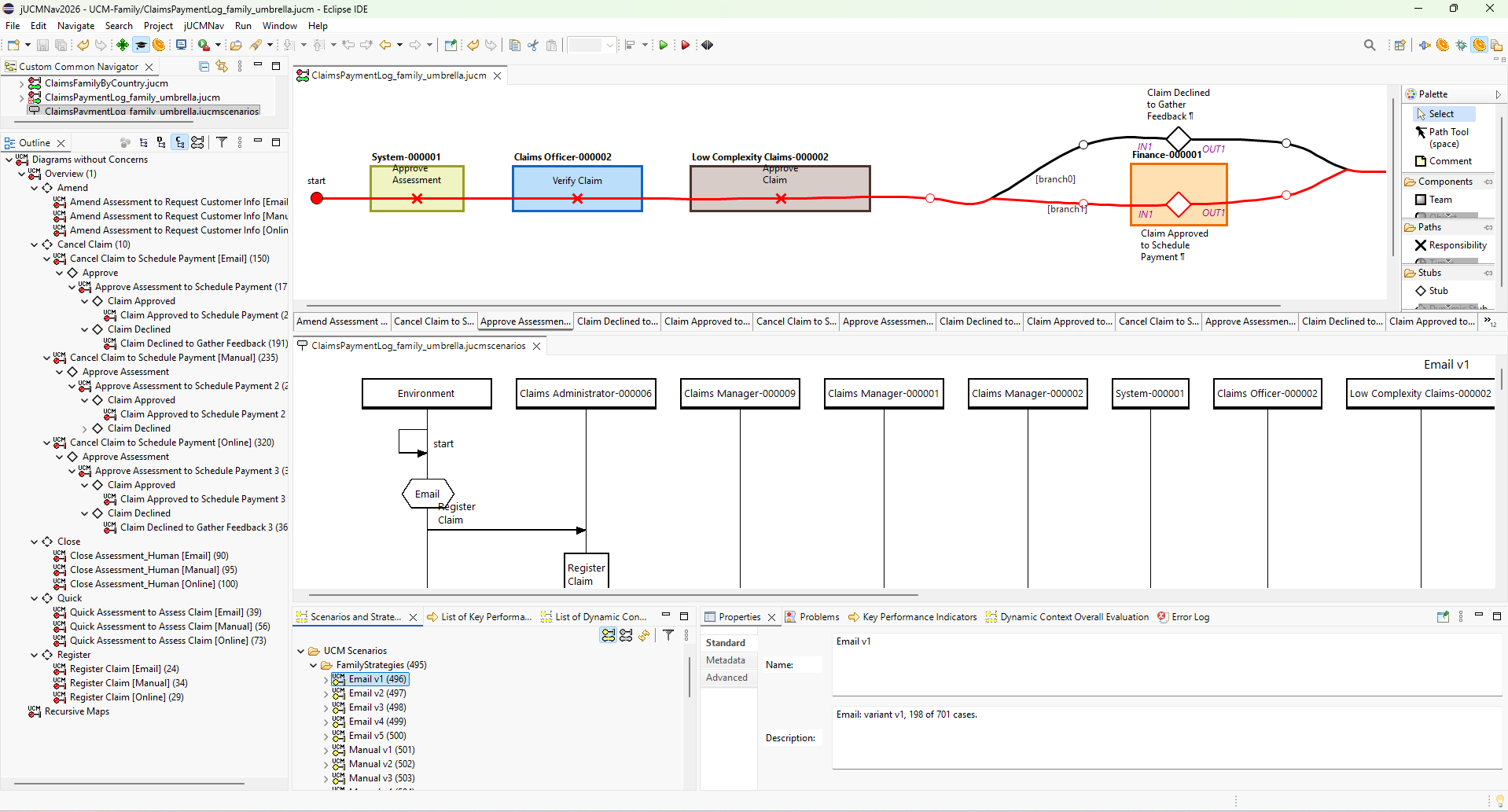}
  \caption{\jUCMNav: scenario execution of a mined UCM and generated sequence diagram.}
  \label{fig:jucmnav}
\end{figure}

\begin{figure}[t]\centering
  \includegraphics[width=\linewidth]{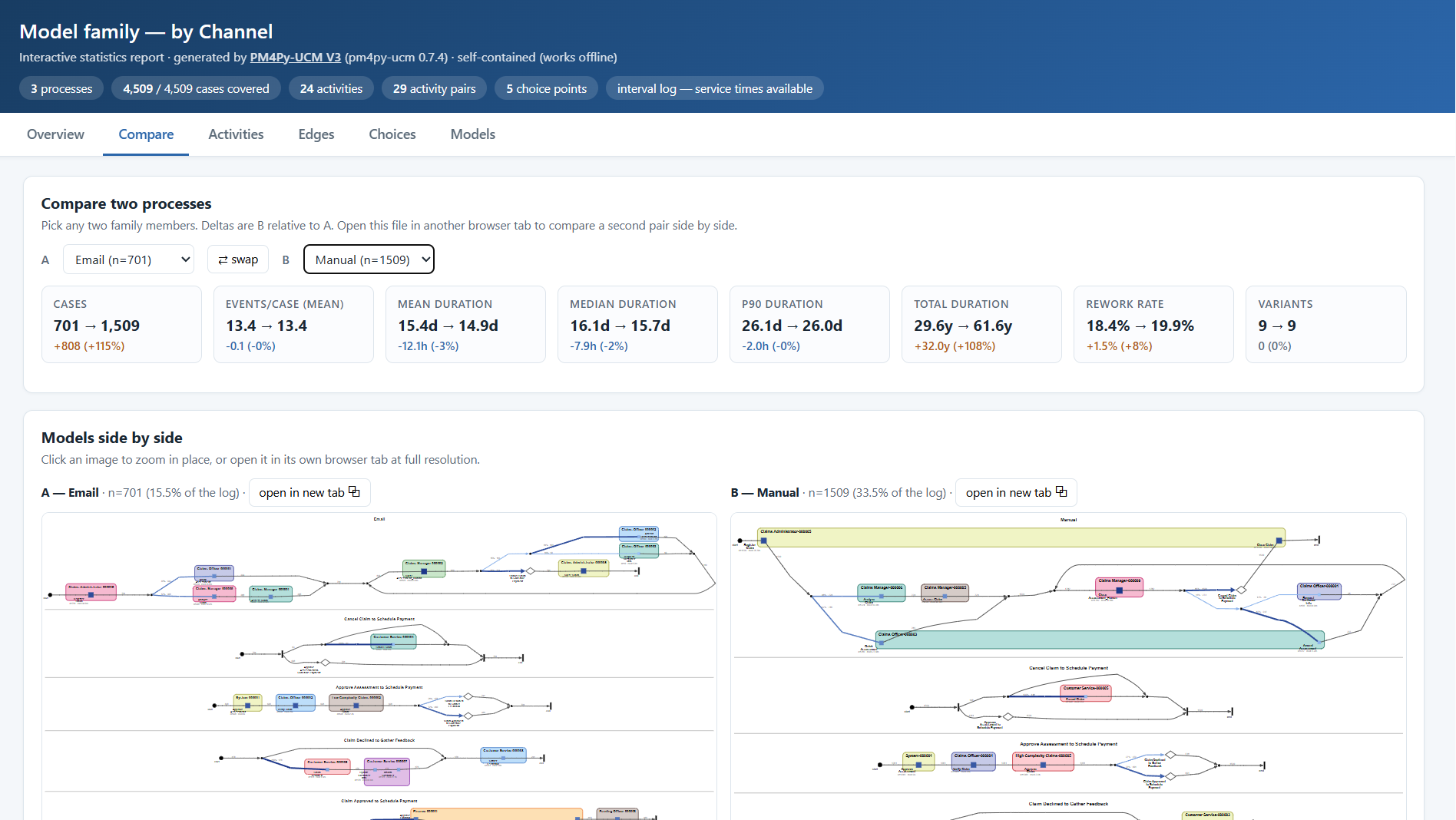}
  \caption{Generated HTML side-by-side comparison report for a UCM model family.}
  \label{fig:compare}
\end{figure}

At the tool's core is a Python object model implementing the UCM subset of the URN metamodel (Z.151), later extended with a \textit{scenario layer} (also from Z.151) to make mined models executable~\cite{pm4pyucmsam}. Discovery is a mainly a chain of model transformations: process tree $\rightarrow$ UCM object model $\rightarrow$  \texttt{.jucm} file. These transformations are where correctness is hardest \textit{and} least visible: a misplaced condition or label yields a \texttt{.jucm} file that opens cleanly in \jUCMNav{} yet executes the wrong scenario, e.g., a silent semantic corruption that raises no exception.

We distinguish here capabilities expected of a mature process-mining tool~\cite{loyola2023PM} from those that are, to our knowledge, distinctive; several of the latter were introduced in prior work~\cite{pm4pyucmrew,pm4pyucmsam} while others are \newbadge{} (\contrib{C1}). Some of the \textit{expected} capabilities of \PUCM{} include:
\begin{itemize}
  \item \newbadge{} Log filtering by specific activities, activity frequency, variant frequency, date ranges (fully inside or intersecting), duration percentile, and top variants by case-coverage \%, with export of the filtered log.
  \item \newbadge{} Activity renaming, with import/export of the renaming map.
  \item \newbadge{} Project/session saving and resuming.
  \item \newbadge{} Performance-metric overlays for activities and branches, and visual metric heat-maps on the process models (Fig.~\ref{fig:bpmn}).
  \item \newbadge{} Dashboard creation with multiple representations, including model pinning (Fig.~\ref{fig:dashboard}).
\end{itemize}

\noindent Among \PUCM's \textit{distinctive} capabilities, we find:
\begin{itemize}
  \item \textit{Log-driven UCM discovery}: to our knowledge, \PUCM{} offers the first pipeline to make Use Case Maps a first-class process-mining output~\cite{pm4pyucmrew}.
  \item \textit{Performer-aware component binding}: activities bound to a two-dimensional hierarchy of components mined from log roles/resources---the ``who'' similar to BPMN pools/lanes, which are seldom mined~\cite{pm4pyucmrew}.
  \item \textit{Hierarchical decomposition} of models using UCM stubs containing sub-processes, to ease the understanding of complex, spaghetti-like models~\cite{pm4pyucmrew}.
  \item \textit{Executable scenario synthesis}: one UCM scenario per concurrency-aware variant, with variant-driven or data-driven (decision-tree~\cite{sklearn2011}) branch conditions~\cite{pm4pyucmsam}. These can be executed for visualization and sequence diagram generation in \jUCMNav{} (Fig.~\ref{fig:jucmnav}).
  \item \newbadge{} \textit{Model families with dynamic-stub variation points}: partitioning a log by 1--2 case attributes yields a family that assembles into a configurable umbrella reference model (and a corresponding \texttt{.jucm} file).
  \item \newbadge{} Family comparisons and dashboards exportable as \textit{single, self-contained interactive HTML files} that embed scalable SVG process models (UCM and BPMN) with navigation hyperlinks to sub-processes (Fig.~\ref{fig:compare}).
  \item \newbadge{} Performance overlays \textit{computed and scaled at three levels}: local (one map), per-family-member (across a family member's decomposed maps, and global (the whole model).
  \item \newbadge{} A \textit{scriptable Python API}: every capability is available programmatically, with no dependence on the Web app. The Web app also supports exporting a whole analysis as \textit{executable Python code}: as a standalone pipeline program and as a JupyterLab Notebook\footnote{\url{https://jupyter.org/}} that doubles as an interactive API tutorial.
\end{itemize}

\section{Reflexive-Mining Method}\label{sec:method}
\begin{figure}
    \centering
    \includegraphics[width=0.875\linewidth]{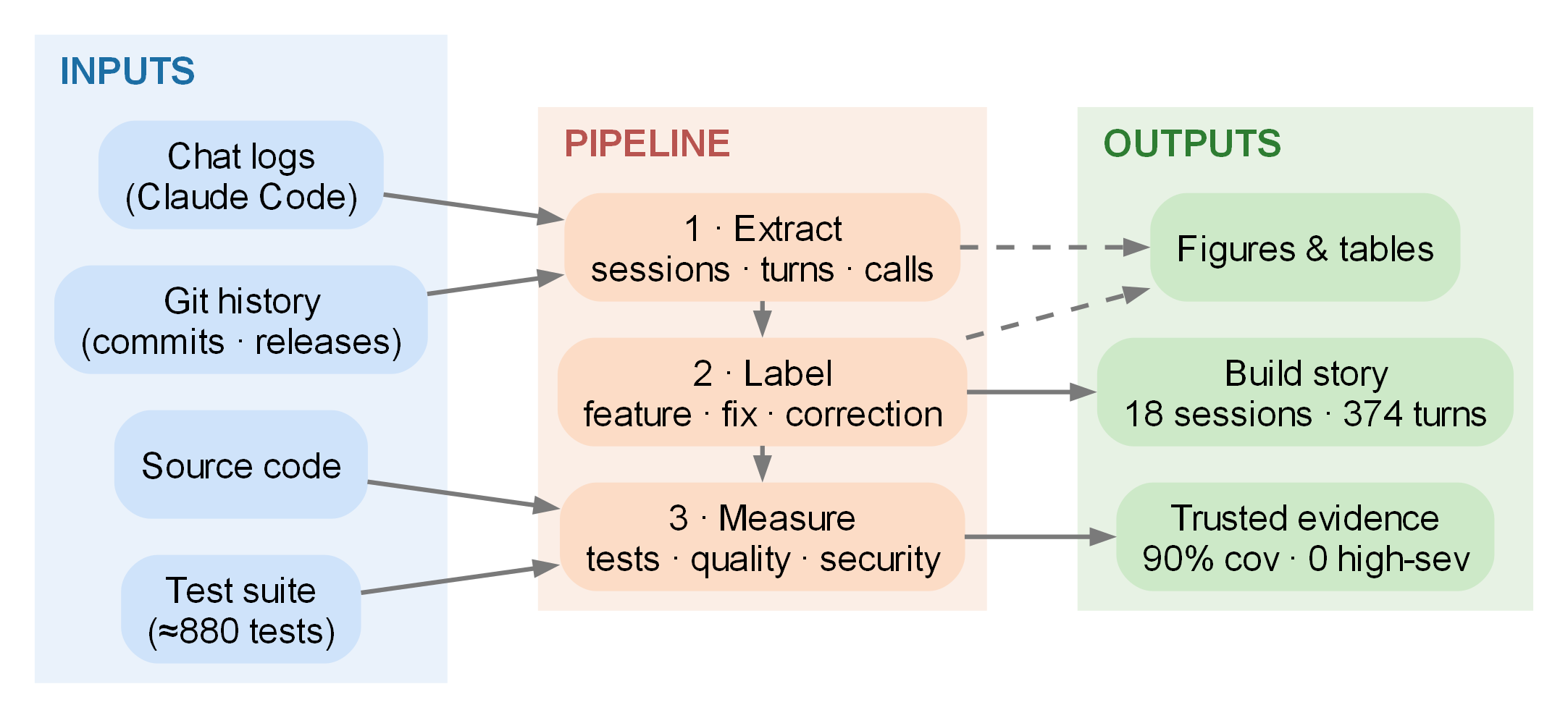}
    \caption{Pipeline applied to \PUCM: From an agent's logs to trustworthy evidence.}
    \label{fig:pipeline}
\end{figure}

The method used to answer the research question is based on the following pipeline (Fig.~\ref{fig:pipeline}), itself implemented in Python (\contrib{C4}).

\textbf{I)~Data Inputs.} 
We draw on three artifacts of the development itself: a)~\textit{18 AI-agent session transcripts} (Claude Code JSONL chat logs with timestamped human turns, agent tool calls, model identifiers, and token usage), b)~the project's \textit{version history} (151 commits, 20 release tags), and c)~the \textit{test suite} tracked across releases. The corpus is frozen at release v0.7.4 (July 22, 2026).

\textbf{II)~Inclusion.}
Some sessions also resume-copy earlier turns verbatim; we collapse 48 such duplicates to their first occurrence. Removing interrupt (e.g., caused by network loss) and attachment artifacts (some elements are in both categories) leaves \textit{317 substantive human turns}, i.e., requests to Claude Code.

\textbf{III)~Taxonomy.}
Each turn receives one \textit{primary intent} and zero or more \textit{cross-cutting flags} (Table~\ref{tab:taxonomy}). In particular, the \texttt{cross\_cutting} flag itself captures requests to propagate a change across auxiliary artifacts (e.g., save/load schema) so they stay consistent after a feature lands.

\begin{table}[h]
  \label{tab:taxonomy}
  \caption{Interaction taxonomy applied to each human turn: exactly one primary intent, plus zero or more (non-exclusive) cross-cutting flags.}
  \centering
  \scriptsize
  \setlength{\tabcolsep}{5pt}
  \renewcommand{\arraystretch}{1.25}
  \rowcolors{2}{white}{MyPurple!15}
  \begin{tabularx}{\linewidth}{@{}XX@{}}
    \toprule
    \multicolumn{2}{@{}l@{}}{\textbf{Primary intent} (chooses exactly one)}\\
    \midrule
    \tax{feature}{add new capability / field / view} &
    \tax{docs}{README, docstrings, tutorials, diagrams, papers}\\
    \tax{fix}{report or repair a defect / wrong behavior} &
    \tax{release\_ops}{versioning, packaging, PyPI, deploy, CI, git operations}\\
    \tax{refactor}{restructure / simplify / rename, no behavior change} &
    \tax{data\_analysis}{running the tool, mining logs, inspecting outputs, notebooks}\\
    \tax{test\_qa}{tests, coverage, oracles, test plan, quality assurance} &
    \tax{clarification}{short answer / approval / redirect (``yes'', ``continue'')}\\
    \tax{design}{options / architecture / planning discussion (pre-code)} &
    \tax{meta\_process}{memory, session setup, plan housekeeping}\\
    \midrule
    \multicolumn{2}{@{}l@{}}{\textbf{Cross-cutting flags} (zero or more)}\\
    \midrule
    \tax{user\_incorrect}{the user's own request/assumption was later shown wrong,
      or retracted (``you're right'', ``never mind'')} &
    \tax{cross\_cutting}{propagate a change across auxiliary artefacts for consistency
      (README, test counts, notebooks, save/load schema, exporters, class diagram)}\\
    \tax{scope\_change}{changes an earlier requirement (``instead'', ``no longer'',
      ``drop'')} &
    \tax{corrects\_agent}{correcting an agent mistake, hallucination, or regression}\\
    \tax{rework}{revisiting work previously considered done} & \\
    \bottomrule
  \end{tabularx}
\end{table}

\textbf{IV) Labeling.} 
Because flags such as \texttt{corrects\_agent} depend on what surrounds a turn, every turn is labeled \emph{in context}. All 374 turns were first labeled by an LLM reading each turn with its in-session neighbors, and a random sample was reviewed by the author; this is the gold standard used throughout. A separate LLM-assisted classifier (cheap \texttt{claude-haiku-4-5} model) reproduces this labeling, and a transparent keyword baseline is retained for auditability (Sect.~\ref{sec:reliability}). The prompt used is as follows (TAXONOMY is from Table~\ref{tab:taxonomy}):
\begin{tcolorbox}[colback=yellow!5!white,colframe=yellow!40!black,left=2pt,right=2pt,top=0pt,bottom=0pt,halign=center]
\begin{footnotesize}
\textsf{You classify a single USER turn from a software-development chat log. Use the\\taxonomy exactly. Consider the provided previous/next turns only as context for the flags (especially user\_incorrect and corrects\_agent). Return STRICT JSON: \\\{"primary":"...","flags":[...], "confidence":0.0-1.0, "rationale":"\textless=15 words"\}.\textbackslash n" + TAXONOMY}
\end{footnotesize}
\end{tcolorbox}

\textbf{V)~Outputs and Privacy.}
We report workflow measures (sessions, tool-action-to-turn ratio, active time as a conservative floor of inter-event gaps $\le$10\,min, output tokens as an effort proxy), commit/churn and release cadence, test growth, and a keyword mapping of turns to modeling concerns. Only \textit{prompt-free, aggregated} results are published; the toolkit lives in a repository\footnote{\url{https://github.com/damyot/pm4py-ucm-devmining}} \textit{separate} from the measured project to avoid an observer effect, and is archived for reproducibility~\cite{zenodo}. The gold labels and all aggregates are released so the coding can be independently audited and re-labeled. The corpus is single-subject (the author's own development), uses no third-party data, and is shared only as aggregates.

\section{Quantitative Development Assessment}\label{sec:results}
\begin{figure}[t]
  \centering
  \includegraphics[width=\linewidth]{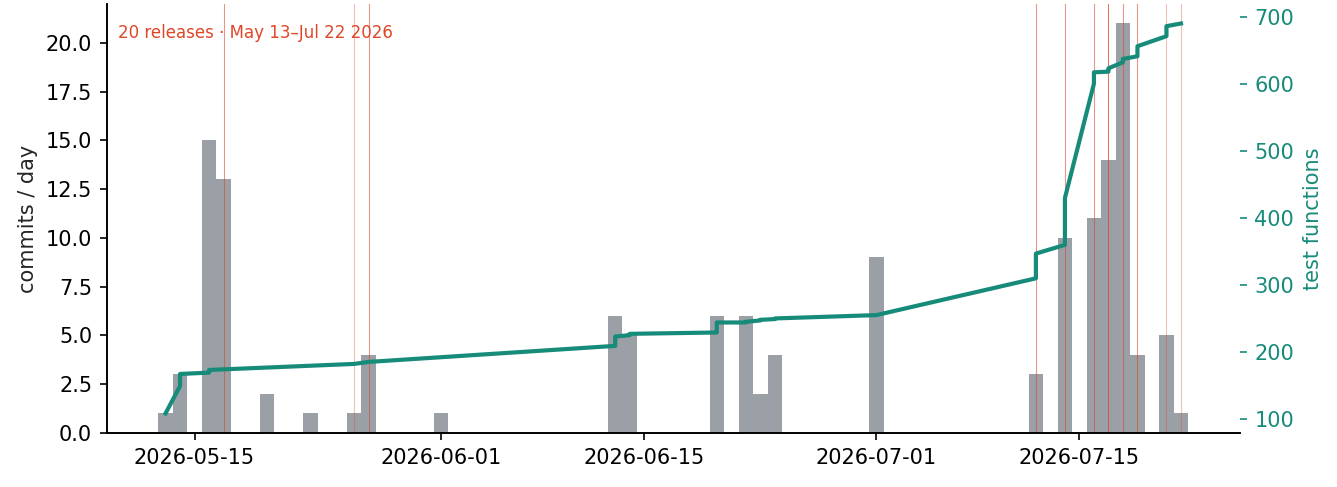}
  \caption{\PUCM{} development timeline: commits/day, releases (orange lines), and test-suite growth.}
  \label{fig:timeline}
\end{figure}

We reconstructed the development of \PUCM{} from 18 AI-agent sessions (2026-05-13\,$\rightarrow$\,07-23, ${\approx}$10 weeks; some sessions spanned more than one day) and the git history, including test cases. All figures below describe this single project (\contrib{C2}): we report counts and proportions of its turns, not inferential statistics. After removing duplicate turns and interrupt/attachment artifacts, \textit{317 substantive human turns} remain. For the original 374 turns, \textit{10{,}328 agent tool actions} were observed, with a \result{${\approx}$28:1 tool-action-to-turn ratio}. Development spanned three model generations, ${\approx}$65\,h of active time, 151 commits ($+75{,}204/-9{,}187$ lines), and 20 releases (v0.2.0\,$\rightarrow$\,v0.7.4). Figure~\ref{fig:timeline} summarizes this timeline.

\subsection{Interaction Mix}
\label{sec:mix}
Table~\ref{tab:intent} 
provides the distributions. \result{Fixes dominate (33.8\%), outnumbering features (14.8\%) at a 2.3:1 ratio}. In essence, the work was overwhelmingly iterative refinement rather than first-shot generation.

\begin{table}
\caption{Primary-intent and cross-cutting-flag distribution over 317 turns.}
\label{tab:intent}
\centering
\scriptsize
\rowcolors{2}{white}{MyPurple!15}
\begin{tabular}{@{}lr r@{\hskip 2em}lr r@{}}
\toprule
\textbf{Primary intent} & \textbf{n} & \textbf{\%} & \textbf{Flag (non-excl.)} & \textbf{n} & \textbf{\%}\\
\midrule
fix           & 107 & 33.8 & corrects\_agent  & 57 & 18.0\\
clarification &  51 & 16.1 & cross\_cutting   & 31 &  9.8\\
feature       &  47 & 14.8 & rework           &  8 &  2.5\\
release\_ops  &  34 & 10.7 & scope\_change    &  4 &  1.3\\
docs          &  26 &  8.2 & user\_incorrect  &  3 &  0.9\\
design        &  24 &  7.6 & & &\\
meta\_process &  21 &  6.6 & & &\\
data\_analysis&   4 &  1.3 & & &\\
refactor      &   2 &  0.6 & & &\\
test\_qa      &   1 &  0.3 & & &\\
\bottomrule
\end{tabular}
\end{table}

\begin{figure}
  \centering
  \includegraphics[width=\linewidth]{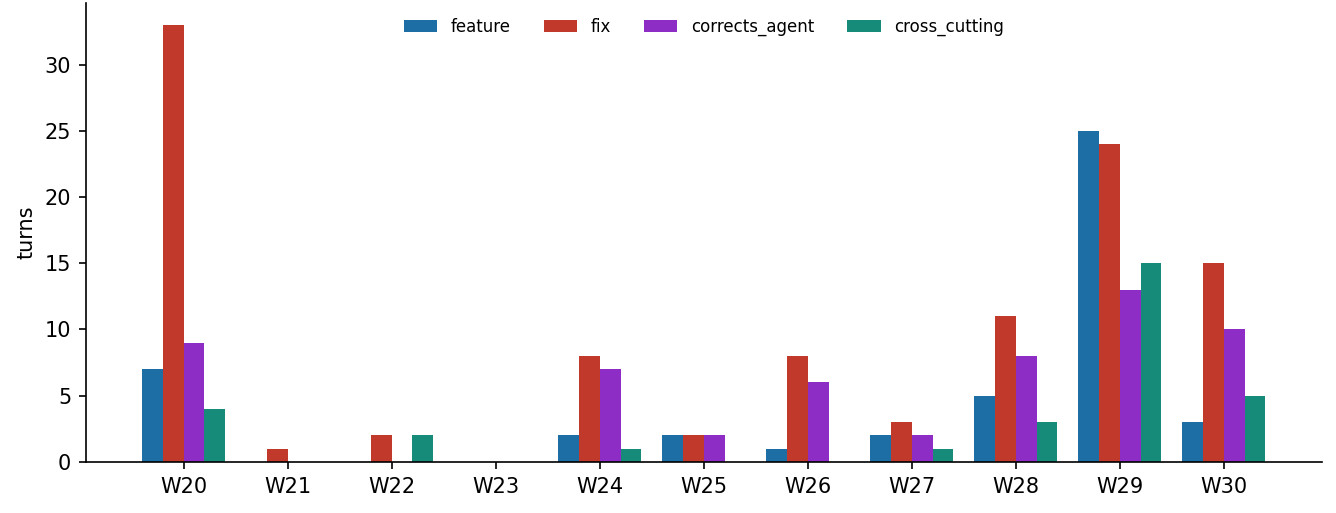}
  \caption{Weekly interaction mix. Bursts: core W20 (2026) \& dashboards/families W29.}
  \label{fig:weekly}
\end{figure}

In terms of correction and cross-cutting effort, \texttt{corrects\_agent} fires on \result{18.0\% of turns (57/317; nearly one turn in five repaired an agent mistake)}, whereas \texttt{user\_incorrect} occurs in only ${\approx}1\%$ of turns (three identified instances): friction came overwhelmingly from agent error rather than human error. This is a conservative lower bound; the flag requires an explicit in-transcript retraction. The developer-proposed \texttt{cross\_cutting} flag (9.8\%) captures consistency work, peaking with the dashboards/families wave in week~29 (Fig.~\ref{fig:weekly}).

\subsection{Labeling Reliability}
\label{sec:reliability}
We initially attempted a transparent, keyword-based classifier as a cheap baseline. The context-aware LLM-based labels, partially validated by the domain expert, remain our reference gold standard. Hence, the comparison measures the \textit{baseline's} inadequacy, not the reference's reliability. The baseline agrees poorly and, crucially, fails to detect agent corrections at all, in part because correcting the agent is a semantic act, not a lexical one. The baseline reliably captured only lexically-marked categories such as cross-cutting consistency work. We therefore \result{rely on the context-aware gold labels} throughout. The full comparison, including per-class precision/recall and Cohen's $\kappa$, is reported in the online appendix~\cite{zenodo}.

\subsection{Effort by Modeling Concern}
\label{sec:topics}
Restricting to the 245 topic-bearing turns (further excluding the \texttt{clarification} and \texttt{meta\_process} categories), effort concentrated on scenarios \& conditions and on visualization \& layout (Table~\ref{tab:topics}). The cross-tabulation reveals that visual/layout work is 78\% fixes and \jUCMNav{} export 83\% fixes. \result{The coding agent struggled most with external-format and visual fidelity} (XMI structure, PNG/SVG rendering, exact condition semantics). Agent corrections cluster on scenarios/conditions (16) and visualization (10), whereas cross-cutting consistency work concentrates on documentation (8).

\begin{table}[t]
\caption{Effort by modeling concern $\times$ intent (10 top concerns, 199/245 topic-bearing turns).
\#Corr = agent-correction turns; \#CC = cross-cutting turns.}
\label{tab:topics}
\centering
\scriptsize
\rowcolors{2}{white}{MyPurple!15}
\begin{tabular}{@{}lccccccc@{}}
\toprule
\textbf{Concern} & \textbf{N} & \textbf{\#Fixes} & \textbf{\#Features} & \textbf{\%Fix} & \textbf{\#Corr} & \textbf{\#CC}\\
\midrule
scenarios \& conditions      & 38 & 18 & 8 & 47 & 16 & 4\\
visualization \& layout      & 36 & 28 & 8 & 78 & 10 & 2\\
mining \& filtering          & 22 & 11 & 6 & 50 &  2 & 1\\
release \& CI                & 21 &  3 & 0 & 14 &  0 & 4\\
dashboards                   & 18 & 10 & 3 & 56 &  5 & 1\\
web/Streamlit UI             & 17 &  8 & 4 & 47 &  5 & 2\\
docs                         & 15 &  1 & 0 &  7 &  0 & 8\\
families \& compare          & 13 &  9 & 3 & 69 &  6 & 0\\
performance \& heat-maps     & 13 &  4 & 5 & 31 &  2 & 3\\
\jUCMNav{} export/round-trip    &  6 &  5 & 0 & 83 &  5 & 0\\
\bottomrule
\end{tabular}
\end{table}


\subsection{Quality Kept Pace with Features}
\label{sec:quality-pace}
\begin{figure}[t]\centering
  \includegraphics[width=\linewidth]{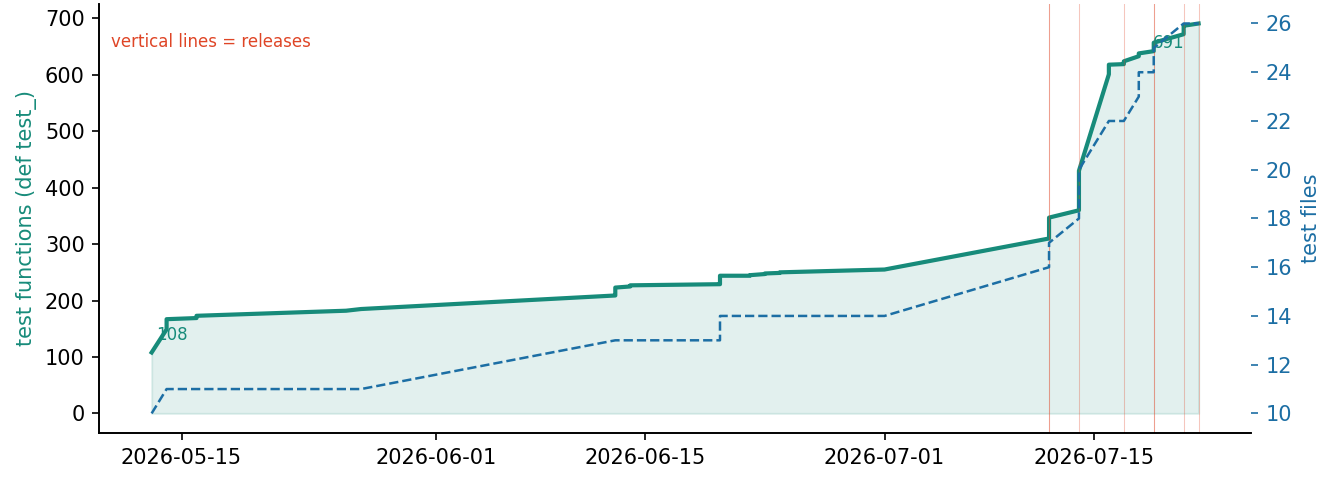}
  \caption{Test-suite growth over the full commit history: 108 test functions on 2026-05-13 to 691 on 2026-07-22 (test files on the right axis; releases marked).}
  \label{fig:tests}
\end{figure}

The test suite grew from 108 test functions (v0.2.0) to 691 by v0.7.4 (10$\rightarrow$26 test files; parametrization expands these into about 880 actual test cases executed). The test numbers rose in lockstep with feature waves (Fig.~\ref{fig:tests}), indicating that tests were written right away, not retrofitted.  Test-line insertions are ${\approx}17\%$ of all insertions. Crucially, the suite does more than exercise code: \result{it closes the ``the agent said it works'' gap for the transformation chain} through \textit{oracle-based} checks (hand-computed fixtures), \textit{algebraic invariants} and \textit{metamorphic} transforms~\cite{metamorphic2018} on the mined models, \textit{byte-deterministic} XMI round-trips, and differential comparison against \PMPy's own metric functions (e.g., about durations). Such testing was also enabled by embedding sample event logs in the tool and its repository. This validation strategy, not the agent's assurances or claims, is what helped make this agent-written modeling tool trustworthy, and the 18\% correction rate shows why it was necessary.

\subsection{Independent Quality, Security, and Architecture}
\label{sec:quality}
Beyond our own tests, we ran standard third-party analyzers on the agent-written code, so the trust argument does not rest on our oracles/tests alone (Table~\ref{tab:quality}, \contrib{C3}). The library is both well covered and well structured, as assessed by \textsf{pytest-cov}\footnote{\url{https://pypi.org/project/pytest-cov/}}: \result{90.6\% line coverage} (7{,}236 of 7{,}983 lines over ${\approx}880$ pytest cases). Static metrics computed with \textsf{radon}\footnote{\url{https://github.com/rubik/radon}} give an average \result{cyclomatic complexity of 5.6} (69\% of blocks rated A; worst case 47, in the log-partitioning routine) and a \result{mean maintainability index of 70} (95\% of files rated A).

Static security analysis with \textsf{Bandit}\footnote{\url{https://github.com/pycqa/bandit}} over ${\approx}25$k LOC (${\approx}19.9$k Source LOC) reports \result{no medium- or high-severity findings}; the twelve low-severity flags are benign (enumeration string literals misread as passwords, \texttt{shell=False} sub-process calls, intentional fallback \texttt{except} blocks). 

\begin{figure}[t]\centering
  \includegraphics[width=\linewidth]{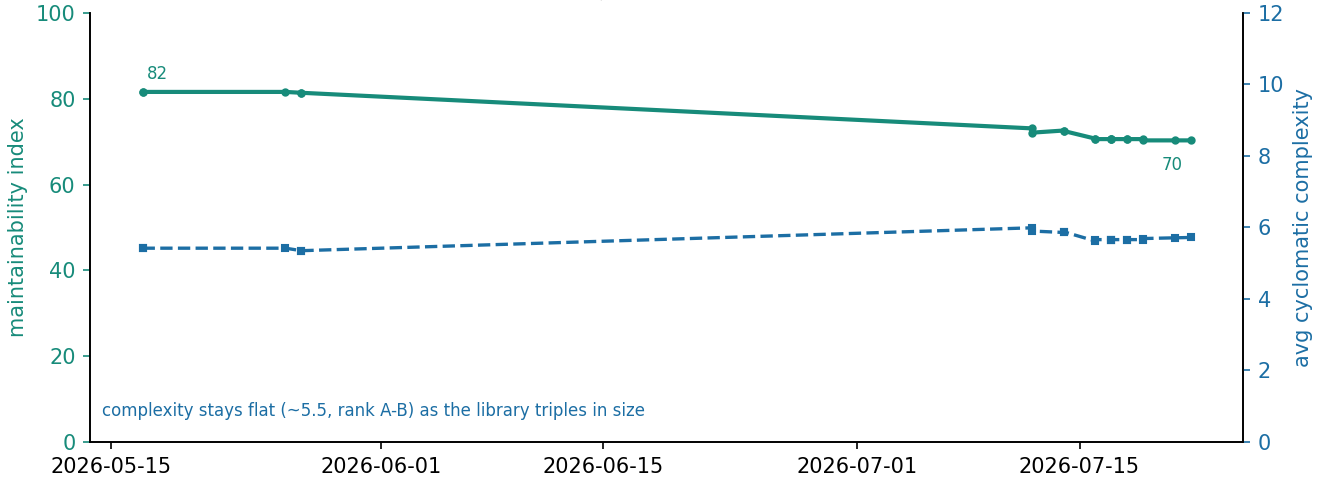}
  \caption{Library maintainability index (left) and average cyclomatic complexity (right) across releases (v0.2.0 to v0.7.4): complexity stays flat (${\approx}5.5$, rank A--B) while maintainability declines only mildly as the code triples in size.}
  \label{fig:quality}
\end{figure}

Architectural contracts checked with \textsf{import-linter}\footnote{\url{https://github.com/seddonym/import-linter}} hold: the object model (metamodel) does not import the discovery algorithms, and reaches the visualization layer only through a single documented layouter reuse. A dependency graph over the 64 code module files shows \result{low coupling} (71 inter-module imports). Figure~\ref{fig:quality} tracks the trend: as the library tripled in size, average complexity stayed flat at ${\approx}5.5$ (rank A--B) while the maintainability index declined only mildly (82$\rightarrow$70, remaining maintainable).

Dependency-vulnerability scanning (\textsf{pip-audit}\footnote{\url{https://github.com/pypa/pip-audit}}) was run against the OSV data\-base\footnote{A distributed vulnerability database for Open Source: \url{https://osv.dev/}} and reported \result{no known vulnerabilities} across the resolved dependency closure (59 packages). The declared version lower bounds of \textsf{Streamlit} and \textsf{scikit-learn} admit older advisories, but the latest versions cause no warnings.


\begin{table}[t]
\caption{Independent quality, security, and architecture metrics for the \PUCM{} library (from third-party analyzers).}
\label{tab:quality}
\centering
\scriptsize
\rowcolors{2}{white}{MyPurple!15}
\begin{tabular}{@{}lll@{}}
\toprule
\textbf{Aspect} & \textbf{Metric} & \textbf{Value}\\
\midrule
Tests (\textsf{pytest-cov}) & line coverage / cases & 90.6\% / ${\approx}880$\\
Complexity (\textsf{radon}) & avg (final rel.) / \% rank A / max & 5.6 / 69\% / 47\\
Maintainability (\textsf{radon}) & mean MI / \% rank A / trend & 70 / 95\% / 82$\rightarrow$70 (CC flat)\\
Size (\textsf{radon}) & SLOC / comment ratio & 19.9k / 15\%\\
Security (\textsf{Bandit}) & high / medium / low & 0 / 0 / 12 (benign)\\
Architecture (\textsf{import-linter}) & contracts / coupling & 2/2 hold / 64 mod., 71 imports\\
Dependencies (\textsf{pip-audit}) & known CVEs / packages (OSV) & 0 / 59\\
\bottomrule
\end{tabular}
\end{table}

Overall, \result{quality was controlled, not sacrificed, under rapid agent-driven code and feature growth}. These additional checks also corroborate our internal oracle-based validation with independent, reproducible evidence.

\section{Discussion: Lessons for AI-Assisted PM/EM Tooling}\label{sec:discussion}

Here are important lessons learned while using Claude Code for developing and then analyzing \PUCM{}, which will hopefully shed some light on agentic AI-assisted PM/EM tool building (\contrib{C4}).

\begin{enumerate}
    \item \lesson{Building a PM/EM tool with an agent is a refinement loop}, not one-shot generation. Despite the author's three decades of experience in requirements modeling (including the standardization of URN and the development of \jUCMNav) and one decade in process mining, fixes outnumbered features 2.3:1, and the ${\approx}$28:1 autonomy ratio is accompanied by an 18\% correction rate: the human's dominant activity is \textit{steering and repairing}. Developers should budget for iteration and for a human-in-the-loop who can recognize defects, many of which could be subtle.

    \item \lesson{Agents are weakest where PM/EM tools are most demanding.} Corrections concentrate on (executable) scenarios/conditions and visualization, and the highest fix rates fall on visualization/layout (78\%) and \jUCMNav/XMI export (83\%). LLMs produce \textit{plausible} structure but struggle with \textit{exact} conformance to a standardized metamodel, a reference editor's format (e.g., \jUCMNav's), and a renderer's expectations (e.g., PNG resolution quality). The response is to make these boundaries mechanically checkable with round-trip determinism, golden-file oracles (e.g., valid and representative \texttt{.jucm} files in XMI), visual-diff oracles, and metamodel conformance checks.

    \item \lesson{Model transformations fail silently, so verification must be executable.}~~A logically wrong condition in a UCM model raises no exception; it yields a model that opens cleanly in \jUCMNav{} yet behaves incorrectly. The risk concentrates in the discovery transformation chain and in keeping the Python metamodel faithful to Z.151's as the scenario layer is added. That \texttt{user\_ incorrect} stays near 1\% while \texttt{corrects\_agent} is 18\% shows the human-in-the-loop was the last line of defense. Pairing every feature with an executable oracle~\cite{metamorphic2018} (importing generated models back into \jUCMNav, replaying scenarios, checking invariants) turns silent semantic errors into visible failures. The test suite growing with features (108$\rightarrow$691 functions) is the mechanism that kept an agent-written tool trustworthy.

    \item \lesson{Feature velocity has a consistency tax, and it lands on documentation.} The \texttt{cross\_cutting} flag (9.8\%, concentrated on docs/READMEs) quantifies a familiar software-maintenance pain, amplified when features ship every few days. Agents do this propagation well \textit{when asked} but rarely initiate it; a ``definition of done'' enumerating auxiliary artifacts, handed to the agent as a checklist, would recover much of this cost.

    \item \lesson{Externalised memory and parallel worktrees are practical enablers.} Multi-week agent development spanned resumed Claude Code sessions (limited to 1M tokens) and parallel git worktrees, stitched by explicit memory/hand-off files; this is a reusable practice pattern that scaled.
\end{enumerate}

In this experience report, an under-resourced PM tool reached 20 releases and a five-view analytics (including many new features) application in ten weeks, built largely by an agent. AI agents can lower the barrier to bespoke PM/EM tooling, but only with the verification scaffolding above and a domain expert in the loop. The 18\% of turns that corrected the agent were \textit{expert} corrections about URN and \jUCMNav{} semantics. AI amplified throughput, but did not necessarily replace modeling expertise. Hence, for AI-assisted EM tooling we recommend: 
\begin{tcolorbox}[colback=MyPurple!5!white,colframe=MyPurple!100!white,title=\textbf{Recommendations},left=2pt,right=2pt,top=0pt,bottom=0pt,halign=left]
{\small
1)~Make transformation boundaries mechanically checkable (round-trips, oracles);\\
2)~Pair every feature with an executable oracle;\\
3)~Give the agent a definition-of-done that enumerates cross-cutting artifacts;\\
4)~Involve a domain expert in the loop as the verifier of record.}
\end{tcolorbox}

\section{Threats to Validity}\label{sec:threats}
Here are important threats categorized along three main validity types.

\textbf{Construct.} Labels are interpretive: i)~\texttt{corrects\_agent} may fold in under-specified prompt, and we did not separate genuine agent error from ambiguous instructions, so 18\% is an upper bound on the former; ii)~\texttt{user\_incorrect} is a lower bound as it requires an explicit in-transcript retraction (and some self-caught mistakes were actually removed from the conversations by changing the erroneous prompt itself). Moreover, iii)~the gold labels are LLM-proposed and expert-verified (through a sample) rather than independently human-coded.

\textbf{Internal.} Transcripts capture the chat channel only; out-of-chat reasoning, manual edits, and local testing are not fully captured, so agent effort is an upper bound and human effort an underestimate. The developer and analyst are the same person, and an AI assisted both building and analyzing the PM tool; we mitigate the labeling risk with a published taxonomy, gold labels, a reproducible pipeline, and an independent keyword baseline, but a second human coder remains future work.

\textbf{External.} With a single tool, developer (domain expert), and agent family, one domain (PM), and ten weeks, this is a \textit{case study}, not a controlled experiment. Development also spanned three model generations (Opus 4.7/4.8, Fable 5), so this mix reflects a moving agent rather than a fixed one. What we expect to transfer are the \textit{method}, the \textit{qualitative patterns}, and the \textit{analysis tool}. Whether this generalizes to other EM tools (more complex or on a domain different from PM) and to developers with less domain expertise is for future work.

\section{Related Work}
\label{sec:related}
There are several related areas worth discussing here. 

\textbf{AI-assisted software engineering.} A growing literature studies LLM code assistants and their effect on productivity and
defects~\cite{cui2026effects,llm4seslr,sergeyuk2025-AI-coding}. These works measure task completion or suggestion acceptance on general-purpose code. Our analysis suggests a complementary dimension: silent semantic corruption in model transformations. Bhati~\cite{bhati2026agenticAI} also highlights five open problems for agentic development (evaluation, governance, technical debt, skill redistribution, and the economics of attention), but our paper only partially addresses the first one. 

\textbf{Building PM/EM tools.} LLMs are increasingly used to produce modeling content (e.g., UML~\cite{genaiuml2023} or BPMN~\cite{horner2026BPMN}) or to generate code~\cite{dong2025codegenerationllm}. Instead, we use an agent to \textit{build the modeling tool} and measure that construction, which is an under-studied topic. \PUCM{} itself extends a long line of PM methods~\cite{pm4py2023,pmtoolsslr}, but differs from metamodeling platforms~\cite{adoxx2013,memo2014,fourem2014,emreview} enabling the specialization of languages and analyses by design.

\textbf{Method and verification.} We mine the development record itself (e.g., agent transcripts), as in studies of LLM-generated code in repositories~\cite{Ji2026LLMcode}, but for a human-agent pair rather than a human team. We recast model-transformation testing (oracles, metamorphic relations, round-trip determinism~\cite{metamorphic2018}) as the scaffolding that makes agent-written tooling dependable. 

\section{Conclusion}
\label{sec:conclusion}
The answer to our research question from this case is a qualified \textit{yes}: trust in \PUCM{} came not from the agent but from the scaffolding around it. Making the transformation boundaries mechanically checkable (golden-file oracles, metamorphic and deterministic round-trips against a reference metamodel and editor) and keeping a domain expert as the record verifier are what turned silent semantic corruption into visible failures. The ${\approx}18\%$ of turns spent correcting the agent (with expert corrections about URN and \textsf{jUCMNav} semantics) show why both were necessary. Development was fix-dominated (2.3:1 ratio over features) and its consistency tax landed on documentation, yet quality kept pace: tests grew in lockstep with features and independent analyzers found the resulting code well covered, low in complexity, and free of medium/high-severity issues. 

For the EM community entering the era of AI, the message is practical: \result{agents can sharply lower the cost of custom modeling tooling}, but only when the tool's outputs are made verifiable at the metamodel, exporter, and renderer boundaries, with a domain expert in the loop (the last point being echoed in~\cite{bhati2026agenticAI}. \result{Agentic AI amplified throughput, but it did not replace modeling expertise}. The paper is accompanied by a privacy-preserving replication package~\cite{zenodo} (including code and derived results) so others can reproduce our results and re-run the analysis on their own agent-assisted projects. 

Future work follows two tracks. For \PUCM, we consider the LLM-assisted interpretation and recommendations, and inference of goal models (GRL) traceable to the mined UCM. On the method side, we plan replication on other tools and developers/teams, with automated support for multiple repositories, and the proper generation of event logs that PM tools (e.g., \PUCM) can mine. Although this paper represents only one data point, we hope that replication by others in different contexts will enable generalizability through meta-analysis.

\subsubsection*{Acknowledgments.}
This work is supported by the NSERC Discovery grant \textit{Requirements-Oriented Process Mining}. GenAI (Clau\-de Code with the Fable 5 and Opus 4.7/4.8 models) was used in three roles: i)~as the main implementer of \PUCM, the subject of this report; ii)~to mine and analyze the development record and to produce draft figures/tables; and iii)~to assist in drafting and revising some of the accompanying text. The author reviewed/verified the AI-generated content (computations, results, figures, and text) for correctness.

\bibliographystyle{splncs04}
\bibliography{references}
\end{document}